\documentclass[aps,prb,twocolumn,footinbib,showpacs,showkeys,superscriptaddress]{revtex4-2}
\usepackage{graphicx,graphics}
\usepackage{natbib}
\usepackage{lipsum}
\usepackage{dcolumn}
\usepackage{amsmath,amssymb,amsfonts}
\usepackage{latexsym,verbatim}
\usepackage{bm}
\usepackage{color}
\usepackage{siunitx}
\usepackage[breaklinks=true,colorlinks,citecolor=blue,linkcolor=blue,urlcolor=blue]{hyperref}
\begin{document}
	
\title{Super-Planckian radiative heat transfer between coplanar two-dimensional metals}
\author{Tao Zhu}
\email{zhutao@tiangong.edu.cn; phyzht@outlook.com}
\affiliation{School of Electronic and Information Engineering, Tiangong University, Tianjin 300387, People's Republic of China}
\author{Yong-Mei Zhang}
\affiliation{College of Physics, Nanjing University of Aeronautics and Astronautics, Jiangsu 210016, People's Republic of China}
\author{Jian-Sheng Wang}
\affiliation{Department of Physics, National University of Singapore, Singapore 117551, Republic of Singapore}
\date{\today}

\begin{abstract}
	Using the nonequilibrium Green's function formalism, we propose a general microscopic framework to investigate the radiative heat transfer (RHT) between coplanar objects with a square lattice. We employ the obtained formulas to two-dimensional (2D) metal configurations with a tight-binding model and the Drude model. Our results reveal that the RHT between coplanar 2D metals is significantly larger than black-body radiation in both the near and far fields, leading to a global super-Planckian RHT. As the separation distance increases, the heat flux density exhibits a rapid decrease in the near field, followed by a slower decrease and eventual $1/d$ dependence in the far field, while maintaining a much higher magnitude than black-body radiation. Evanescent waves dominate the heat transfer in the near field, while propagating waves dominate the far field. Surprisingly, the propagating heat flux remains almost constant over a wide range of distances, resulting in a super-Planckian behavior in the far field. The dispersion relation of the spectrum function reveals distinct contributions from propagating and evanescent waves, with possible origins from surface plasmon resonance. These findings provide insights into the unique characteristics of RHT between coplanar 2D metals and highlight the potential for achieving enhanced heat transfer beyond the black-body limit. Our method is applicable to any coplanar objects with square lattices, paves the way for expanded investigations into various lattice geometries.
\end{abstract}
\maketitle

\section{Introduction}

In electrodynamics, the upper limit of energy generated by thermal motion is governed by black-body radiation, with the characteristic frequency spectrum determined solely by the temperature of the body, as per Planck's law. However, recent investigations into new tunneling channels of electromagnetic waves have revealed that radiative heat transfer (RHT) between bodies can exceed the black-body limit, resulting in a phenomenon known as super-Planckian RHT \cite{super1,super2,super3}. The most extensively studied super-Planckian RHT is the well-known near-field RHT \cite{near1,near2,near3,near4}, where the separation distance between two bodies is less than Wien's wavelength. Through tunneling evanescent waves, the heat flux between bodies in the near field can significantly surpass the black-body limit by several orders of magnitude. Owing to its potential applications in a wide range of innovative technologies, such as nanoscale energy harvesting \cite{energy-conversion} and thermal management \cite{thermal-manage}, super-Planckian energy transport has generated tremendous research interest \cite{cuevas,heat,zhu1,principi}.

Initially, super-Planckian RHT was considered present only in the near field, where new tunneling channels such as evanescent waves can exist. In contrast, propagating waves dominate far-field RHT, and Kirchhoff's law governs the heat emission, thus bounded by the black-body limit \cite{biehs1}. However, recent studies have revealed that far-field super-Planckian RHT can be achieved between sub-wavelength objects \cite{Hurtado1,Thompson,Hurtado2}. When the dimensions of the bodies are smaller than the thermal wavelength, the heat flux between them in the far field can also exceed the black-body limit with a defined view factor. For instance, experimental work by Thompson \textit{et al.} demonstrates that far-field RHT between planar membranes with sub-wavelength dimensions can exceed the black-body limit by more than two orders of magnitude \cite{Thompson}. Fern\'andez-Hurtado \textit{et al.} performed further theoretical investigations to explore the limits of super-Planckian far-field RHT using two-dimensional (2D) materials. Their results show that the exchanged thermal radiation between two coplanar graphene flakes can be more than seven orders of magnitude larger than the black-body limit, with the enhancement of RHT in the far-field dominated by TE-polarized guiding modes \cite{Hurtado2}. 

Theoretical works on RHT have generally been based on the fluctuational electrodynamics (FE) theory of Polder and van Hove \cite{pvh}, using Rytov's formulation of fluctuating electromagnetic fields \cite{rytov}. However, previous studies have often relied on macroscopic local response functions for coplanar objects, which may be insufficient to describe RHT in the extreme near field (distances approaching atomic lattice constants) and in materials with significant inhomogeneities where local field effects are non-negligible \cite{zhu2}. This limitation may hinder the application of the macroscopic local model to subwavelength objects, which are indispensable for achieving far-field super-Planckian RHT, and suggests the need for a microscopic nonlocal response function \cite{chapuis,walter}.

Alternatively, a general microscopic nonequilibrium Green's function (NEGF) approach for photon transport, inclusive of RHT, has been developed \cite{wang1,wang2,wang3,zhang}. This approach constructs the thermal transmission function from the photon Green's function coupled with the self-energies of the bodies involved. One advantage of the NEGF formalism is its ability to naturally incorporate the microscopic tight-binding method, enabling the explicit calculation of self-energies through approximations such as the random phase approximation. This microscopic treatment allows for a more accurate description of the electronic structure and the inclusion of quantum effects, which are crucial for understanding RHT at the nanoscale. More importantly, the NEGF framework is completely equivalent to FE theory under conditions of local thermal equilibrium while maintaining the ability to deal with entirely nonequilibrium situations where the fluctuation-dissipation theorem is not applicable \cite{zhang,zhu3,tang,wang5}. This equivalence ensures that the NEGF approach can reproduce the results of FE theory in the appropriate limits \cite{zhu3}, providing a unified description of RHT across different regimes. Despite the potential of the NEGF method for RHT applications, its extensive implementation is still emerging, and a comprehensive quantum-mechanical microscopic method for studying RHT between coplanar objects has yet to be established.

In this work, we harness the NEGF formalism to develop a fully quantum-mechanical microscopic theoretical framework for examining RHT between coplanar objects, which we apply to 2D common metals as an example. To obtain the electronic and response properties of the media, we start from a general tight-binding model with a square lattice and subsequently employ the Drude model that is applicable for pure metals. Our results show that the RHT between coplanar 2D metals exhibits a super-Planckian behavior in both near-field and far-field scenarios, dominated by evanescent and propagating waves, respectively. The calculated heat flux in the near field is found to be about four orders of magnitude higher than the black-body limit and can reach a million-fold enhancement in the far-field. This investigation not only introduces a robust microscopic framework for probing RHT between coplanar 2D objects but also provides a versatile model that can be extended to other metals with appropriate parameters. Furthermore, it lays the groundwork for future explorations of different materials and lattice geometries.

\section{NEGF formalism for RHT between coplanar 2D lattices}

\begin{figure}
	\centering
	\includegraphics[width=8.6 cm]{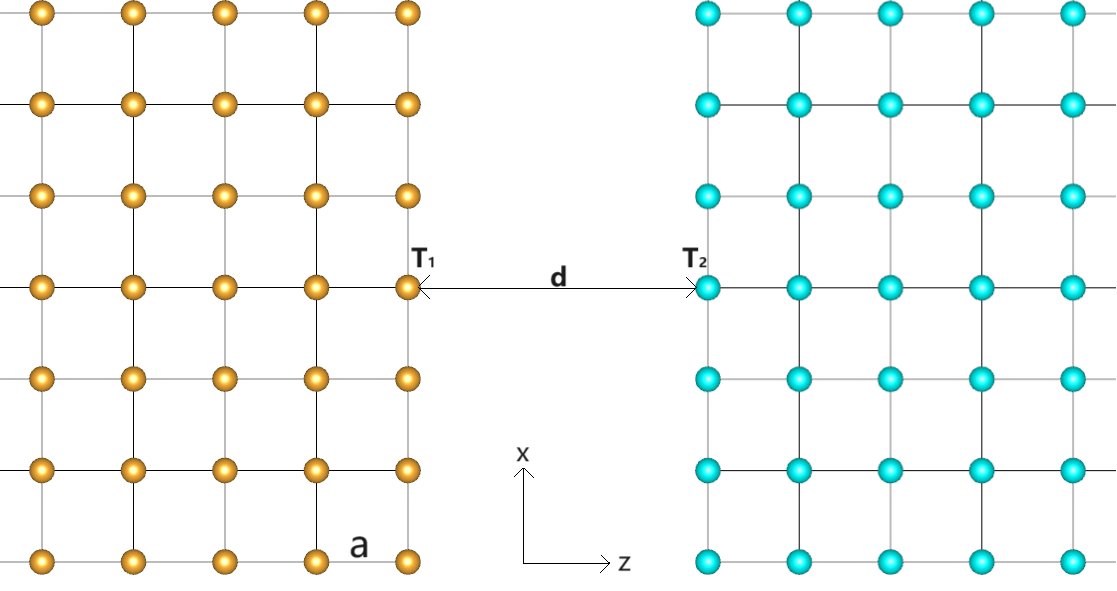}
	\caption{Model of two coplanar 2D objects with a lattice constant $a$ separated by a vacuum gap of $d$. Each lattice is in its internal thermal equilibrium state; lattice 1 is at temperature $T_1$, and lattice 2 is at temperature $T_2$. The $x$-direction is periodic, and the $z$-direction is semi-infinite.}
	\label{fig1}
\end{figure}

We consider two semi-infinite 2D square lattices placed in the $x$--$z$ plane, as shown in Fig.~\ref{fig1}. Both lattices have the same lattice constant $a$ and are separated by a vacuum gap of size $d$. For each lattice, we assume electrons are located at the lattice sites labeled by $l=(l_x,l_z)$ and can only hop to the nearest-neighbor sites within their own lattice, i.e., no electrons can hop from one lattice to another. If the temperatures $T_1$ and $T_2$ of the two lattices are different, radiative heat transfer occurs, and we aim to calculate the net heat flux density between them.

In both the FE and NEGF formalisms of RHT, the energy current between two bodies is given by a Landauer-like formula \cite{kruger1}:
\begin{equation}
	J= \int_{0}^{\infty}\dfrac{d\omega}{2\pi} \hbar\omega \bigl[N_1(\omega)-N_2(\omega)\bigr]T(\omega),
	\label{landauer}
\end{equation}
where $N_\alpha(\omega) = \left[e^{\hbar\omega/(k_BT_\alpha)}-1\right]^{-1}$ is the Bose distribution function at the temperature $T_\alpha$ for object $\alpha$. The transmission function $T(\omega)$ characterizes the coupling between objects mediated by fluctuating electromagnetic fields, which can be expressed using different terminologies in the FE and NEGF methods.

In this work, we adopt the microscopic NEGF formalism. Under the local equilibrium approximation, Eq.(\ref{landauer}) can be derived from the Meir-Wingreen formula \cite{meir1,meir2}, and the transmission coefficient $T(\omega)$ is given by the Caroli formula \cite{zuquan,jiang}:
\begin{equation}
	T(\omega)={\rm Tr}\big[D_{21}^r\Gamma_1 D_{12}^a\Gamma_2\big],
	\label{caroli}
\end{equation}
where the superscripts $r$ and $a$ denote the retarded and advanced components, respectively. The central quantities for calculations in Eq.(\ref{caroli}) are the photon Green's function $D$ and the spectrum function $\Gamma$, which is defined as $\Gamma =i(\Pi^r-\Pi^a)$, where the photon self-energy $\Pi$ describes electron-photon interactions within each object. Given that the advanced components are the conjugate transpose of the retarded ones, i.e., $D^a=(D^r)^\dagger$ and $\Pi^a=(\Pi^r)^\dagger$, we can calculate the heat flux between the two objects using the equations above once the retarded photon Green's function $D^r$ and self-energy $\Pi^r$ of the system are obtained.

To obtain the retarded photon Green's function $D^r$ and self-energy $\Pi^r$, we consider a vector potential ${\bf A}$ and its interaction with electrons within a tight-binding model framework \cite{wang6}. As gauge invariance uniquely determines the form of interactions between electrons and fields, by adopting the temporal gauge (where the scalar potential $\phi = 0$), the Hamiltonian of the interacting system can be written as \cite{wang1}:
\begin{eqnarray}
	\nonumber\hat{H} &=&\dfrac{\epsilon_0}{2}\int dV \biggl[ \biggl(\dfrac{\partial {\bf A}}{\partial t}\biggr)^2 + c^2(\nabla\times {\bf A})^2\biggr]\\
	&+&\sum_{l,l^\prime}c^\dagger_lH_{l,l^\prime}c_{l^\prime} {\exp}\biggl(\dfrac{e}{i\hbar}\int_{l^\prime}^{l} {\bf A} \cdot d {\bf l} \biggr),
	\label{hamiltonian}
\end{eqnarray}
where $l$ denotes the electron sites in the 2D lattice, $H_{l,l^\prime}$ is the single-electron Hamiltonian matrix element, $c_l$ and $c^\dagger_l$ represent the annihilation and creation operators applied on site $l$, respectively. $\epsilon_0$ is the vacuum permittivity and $c$ is the speed of light. The integral in the exponential function is a line integral from site $l^\prime$ to site $l$ following a straight path.

We now invoke the NEGF machinery, in which the contour-ordered photon Green's function $D$ and photon self-energy $\Pi$ are defined as \cite{wang5,datta}:
\begin{eqnarray}
	&&D_{\mu\nu}({\bf r}\tau;{\bf r}^\prime\tau^\prime)=\dfrac{1}{i\hbar}\bigl\langle T_c A_\mu({\bf r},\tau)A_\nu({\bf r^\prime},\tau^\prime)\bigr\rangle,\\
	&&\Pi_{l\mu;l^\prime\nu}(\tau;\tau^\prime)=\dfrac{1}{i\hbar}\bigl\langle T_c I_{l\mu}(\tau)I_{l^\prime\nu}(\tau^\prime)\bigr\rangle,
\end{eqnarray}
where $\tau$ and $\tau^\prime$ are Keldysh contour times, $T_c$ is the time-ordering operator on the contour, $\mu$ and $\nu$ represent the $x$ or $z$ directions, and the average is taken over a nonequilibrium steady state. Here, $D$ is defined in the entire space, while $\Pi$ is restricted to the discrete lattice sites, and the current operator $I$ describes the hopping of electrons between different sites.

For the 2D lattice configuration illustrated in Fig.\ref{fig1}, with periodicity only in the $x$-direction and no electron sites in the $y$-direction, we can perform a Fourier transform of the vector potential $A^\mu$ or electron annihilation operator $c_l$ along the $x$ direction, while maintaining the real space representation in the $z$ direction. Due to this periodicity, the electron Hamiltonian becomes block-diagonal after the Fourier transformation, and the fermion operators defined on lattice sites can be represented in the mixed space as follows:
\begin{equation}
	c_{l_x,l_z}=\dfrac{1}{\sqrt{L}}\sum_{q_x}e^{iq_xl_xa}c(q_x,l_z),
\end{equation}
where $L$ is the number of discrete wavevectors in the $x$-direction and the transverse wavevector $q_x$ takes on the values $q_x=2\pi m/(aL)$ for $m=0,1,...,L-1$. Consequently, the Fourier-transformed version of the photon Green's function in our system is expressed by
\begin{equation}
	D_{\mu\nu}(q_x,z,\tau;z^\prime,\tau^\prime)=\dfrac{1}{i\hbar}\bigl\langle T_c A_\mu(q_x,z,\tau)A_\nu(-q_x,z^\prime,\tau^\prime)\bigr\rangle.
\end{equation}

The self-energies of the photons are determined by expanding the exponential term in the interaction part of the Hamiltonian in Eq. (\ref{hamiltonian}) to the second order in $A^\mu$. The linear term leads to the current-vector potential interaction, and after applying standard diagrammatic techniques, the linear term self-energy in contour time is found to be
\begin{equation}
	\Pi_{\mu\nu}(q_x,l_z,\tau;l_z^\prime,\tau^\prime)=\dfrac{1}{i\hbar L}\bigl\langle T_c I_{\mu}(q_x,l_z,\tau)I_\nu(-q_x,l_z^\prime,\tau^\prime)\bigr\rangle.
\end{equation}
Here, the self-energy $\Pi_{\mu\nu}$ is dependent on the transverse wavevector $q_x$ as well as the lattice positions $l_z$ and $l_z^\prime$ in the $z$-direction. The expectation is taken over the equilibrium state, with the current operators $I_{\mu}$ and $I_\nu$ describing the electron hopping processes. 

As the $x$ direction is periodic, electrons in $l_x$ can hop to $l_x\pm1$. We define the ``velocity" of the electron in the $x$ direction as
\begin{equation}
	v(p_x,p_x^\prime) = \frac{at}{\hbar}\bigl(\sin(p_xa) + \sin(p_x^\prime a)\bigr),
\end{equation}
where $a$ is the lattice constant, $t$ is the hopping parameter, and $p_x^\prime$ and $p_x$ are the initial and final momenta of the electron, respectively. The current operator in the $x$-direction is then given by
\begin{eqnarray}
	\nonumber I_x(q_x,l_z,\tau)=&-&e\sum_{p_x,p_{x^\prime}}v(p_x,p_x^\prime)c^\dagger(p_x,l_z,\tau)\\
	&\times& c(p_x^\prime,l_z,\tau)\delta(p_x^\prime-p_x-q_x),
\end{eqnarray}
where $(-e)$ is the electron charge, $c^\dagger$ and $c$ are the creation and annihilation operators for electrons, and $\delta$ is the Kronecker delta function since our momentum labels are discrete, ensuring momentum conservation.

\begin{widetext}
	The situation differs in the $z$ direction as electrons at the boundary site can only hop to inner sites. A central difference operator, $\Delta c^{(\dagger)}(p_x^\prime,l_z,\tau) = c^{(\dagger)}(p_x,l_z+1,\tau) - c^{(\dagger)}(p_x,l_z-1,\tau)$, is defined to handle this situation, leading to the expression for the $z$-component of the current operators:
	\begin{eqnarray}
		I_z(q_x,l_z,\tau)=\dfrac{ieat}{2\hbar}\sum_{p_x,p_{x^\prime}}\big [c^\dagger(p_x,l_z,\tau)\Delta c(p^\prime_x,l_z,\tau)-\Delta c^\dagger(p_x,l_z,\tau)c(p_x^\prime,l_z,\tau)\big]\delta(p_x^\prime-p_x-q_x).
	\end{eqnarray}
	Using the notation $G_{AB}(\tau,\tau^\prime) = (\frac{1}{i\hbar}) \langle T_c A(\tau) B(\tau^\prime) \rangle$ for the electron Green's function $G$ and applying the Wick theorem \cite{datta}, we can obtain the photon self-energies in the following matrix sectors:
	\begin{eqnarray}
		\label{Pixx}
		\Pi_{xx}^{(1)}(q_x,l_z,\tau;l^\prime_z,\tau^\prime)&=&\dfrac{\hbar e^2}{iL}\sum_{p_x,p_x^\prime}v^2(p_x,p_x^\prime)G(p_x,l_z,\tau;l^\prime_z,\tau^\prime)
		G(p_x^\prime,l^\prime_z,\tau^\prime;l_z,\tau)\delta(p_x^\prime-p_x-q_x),\\
		\label{Pixz}
		\Pi_{xz}^{(1)}(q_x,l_z,\tau;l^\prime_z,\tau^\prime)&=&\dfrac{e^2at}{2L}\sum_{p_x,p_x^\prime}v(p_x,p_x^\prime)\big [G_{c\Delta c^\dagger}(p_x,l_z,\tau;l^\prime_z,\tau^\prime)G(p_x^\prime,l^\prime_z,\tau^\prime;l_z,\tau)\\
		\nonumber&&-G(p_x,l_z,\tau;l^\prime_z,\tau^\prime)G_{\Delta cc^\dagger}(p_x^\prime,l^\prime_z,\tau^\prime;l_z,\tau)\big ]
		\delta(p_x^\prime-p_x-q_x),\\
		\label{Pizx}
		\Pi_{zx}^{(1)}(q_x,l_z,\tau;l^\prime_z,\tau^\prime)&=&\dfrac{e^2at}{2L}\sum_{p_x,p_x^\prime}v(p_x,p_x^\prime)\big [G(p_x,l_z,\tau;l^\prime_z,\tau^\prime)G_{c\Delta c^\dagger}(p_x^\prime,l^\prime_z,\tau^\prime;l_z,\tau)\\
		\nonumber&&-G_{\Delta cc^\dagger}(p_x,l_z,\tau;l^\prime_z,\tau^\prime)G(p_x^\prime,l^\prime_z,\tau^\prime;l_z,\tau)\big ] \delta(p_x^\prime-p_x-q_x).
	\end{eqnarray}
	The most complex $zz$ component is given by
	\begin{eqnarray}
		\label{Pizz}
		\Pi_{zz}^{(1)}(q_x,l_z,\tau;l^\prime_z,\tau^\prime)&=&\dfrac{i(eat)^2}{4\hbar L}\sum_{p_x,p_x^\prime}\Big [G_{\Delta cc^\dagger}(p_x,l_z,\tau;l^\prime_z,\tau^\prime)G_{\Delta cc^\dagger}(p_x^\prime,l^\prime_z,\tau^\prime;l_z,\tau)\\
		\nonumber &&-G(p_x,l_z,\tau;l^\prime_z,\tau^\prime)G_{\Delta c\Delta c^\dagger}(p_x^\prime,l^\prime_z,\tau^\prime;l_z,\tau)-G_{\Delta c\Delta c^\dagger}(p_x,l_z,\tau;l^\prime_z,\tau^\prime)G(p_x^\prime,l^\prime_z,\tau^\prime;l_z,\tau)\\
		\nonumber &&+G_{c\Delta c^\dagger}(p_x,l_z,\tau;l^\prime_z,\tau^\prime)G_{c\Delta c^\dagger}(p_x^\prime,l^\prime_z,\tau^\prime;l_z,\tau)\Big ]\delta(p_x^\prime-p_x-q_x).
	\end{eqnarray}
In the above formulas, if the site index $l_z$ appears to be outside the lattice due to the central difference operator $\Delta$, 
the correponding term is understood to be zero.
	Equations (\ref{Pixx})-(\ref{Pizz}) give the photon self-energy $\Pi$ in contour time. To apply Eq.~(\ref{caroli}), we need to transform the contour time formulas to real-time, which is achieved in the frequency domain for the retarded component by 
the Langreth rule \cite{langreth} as 
	\begin{equation}
		G_1(\tau,\tau^\prime)G_2(\tau^\prime,\tau)\to\int_{-\infty}^{+\infty}\dfrac{dE}{2\pi\hbar}\Big[G^r_1(E)G^<_2(E-\hbar\omega)+ G^<_1(E)G^a_2(E-\hbar\omega)\Big],
	\end{equation}
	where $G^<$ denotes the lesser Green's function.
	
	Next, we consider the quadratic term ($A^2_\mu$) in the expansion, which gives a plasmon or diamagnetic contribution. This term is 
important to maintain gauge invariance.  After tedious derivations, the results are diagonal in direction and site indices. The $xx$ component of the plasmon contribution to the retarded photon self-energy in energy space is given by:
	\begin{eqnarray}
		\label{Pixx2}
		\Pi_{xx}^{r(2)}(q_x,\omega,l_z,l^\prime_z)=\dfrac{e^2}{imL}\delta_{l_z,l^\prime_z}\sum_{p_x}\int_{-\infty}^{+\infty}\dfrac{dE^\prime}{2\pi}{\rm cos}(p_xa)G^<(p_x,E^\prime,l_z,l_z),
	\end{eqnarray}
	where the effective mass is defined by the relation $t=\frac{\hbar^2}{2ma^2}$.
	The $zz$ component is expressed as:
	\begin{eqnarray}
		\label{Pizz2}
		\nonumber \Pi_{zz}^{r(2)}(q_x,\omega,l_z,l^\prime_z)=\dfrac{e^2}{4imL}\delta_{l_z,l^\prime_z}\sum_{p_x}\int_{-\infty}^{+\infty}&&\dfrac{dE^\prime}{2\pi} \Big[ G^<(p_x,E^\prime,l_z,l_z+1)+G^<(p_x,E^\prime,l_z+1,l_z)\\
		&&+G^<(p_x,E^\prime,l_z,l_z-1)+G^<(p_x,E^\prime,l_z-1,l_z) \Big].
	\end{eqnarray}
	The total retarded photon self-energy is then given by the sum of the linear and quadratic contributions, $\Pi^r = \Pi^{r(1)}+\Pi^{r(2)}$.  The above expressions for $\Pi^r$ are known as random phase approximation, as higher-order electron-photon couplings are ignored. 
\end{widetext}

In the scenario where the system is in local thermal equilibrium, meaning that the temperature is well-defined, we can employ the fluctuation-dissipation theorem \cite{fdt1,fdt2}. This theorem allows us to relate the lesser Green's function $G^<$ to the retarded ($G^r$) and advanced ($G^a$) Green's functions as follows:
\begin{equation}
	G^<=-f(G^r-G^a),
\end{equation}
where $f = \left[e^{(E-\mu)/(k_BT)}+1\right]^{-1}$ is the Fermi-Dirac distribution function at temperature $T$ and chemical potential $\mu$.

To derive the retarded electron Green's function $G^r$, we can focus on the right side of the system. Denote $c(q_x)$ as the semi-infinite vector of annihilation operators for layers $1, 2, ..., l_z, ...$. The Hamiltonian for the right system, which is block-diagonal with hopping parameter $t$ and electron dispersion $\epsilon_{1D} = -2t\cos(q_xa)$, allows us to express the free electron Green's function in terms of the inverse of the Hamiltonian:
\begin{equation}
	G^r(q_x,E) = \bigl[E+i\eta-H(q_x)\bigr]^{-1},
\end{equation}
where $H(q_x)$ is the single-particle Hamiltonian as a matrix indexed by the position $l_z$, and $\eta$ is a small damping factor that accounts for electron relaxation processes. An explicit expression for the electron Green's function is:
\begin{equation}
	G^r(q_x,E,l_z,l^\prime_z) = \frac{\lambda^{l_z+l^\prime_z} - \lambda^{|l_z-l^\prime_z|}}{t\left(\frac{1}{\lambda} - \lambda\right)},
\end{equation}
where $\lambda$ is a complex number with $|\lambda| < 1$ that satisfies the quadratic equation:
\begin{equation}
	t + (E + i\eta - \epsilon_{1D})\lambda + t\lambda^2 = 0.
\end{equation}

The subsequent objective is to compute the retarded photon Green's function $D^r$. By Utilizing the standard diagrammatic expansion in the interacting picture, the Dyson equation for the retarded photon Green's function $D^r$ is \cite{wang1,dyson}:
\begin{eqnarray}
	\nonumber D^r_{\mu\nu}&&(q_x,\omega,z,z^\prime)=\\
	\nonumber&&d^r_{\mu\nu}(q_x,\omega,z,z^\prime)
	+\sum_{l_z,l'_z,\alpha,\beta}\Big[d^r_{\mu\alpha}(q_x,\omega,z,al_z)\\
	&&\times\,\Pi^r_{\alpha\beta}(q_x,\omega,l_z,l'_z)D^r_{\beta\nu}(q_x,\omega,al'_z,z^\prime)\Big],
\end{eqnarray}
where $d^r$ represents the free photon Green's function. We only need the solution when $z= a l_z$ or $z'= a l'_z$ on the
electron lattice sites. We can obtain the expression in the frequency domain from the second quantization representation of the vector potential or, alternatively, we can use the equation of motion method. The free retarded photon Green's function is given by \cite{zhang,wang1}:
\begin{equation}
	d^r(\mathbf{q},\omega) = \frac{\mathbf{U} - \mathbf{q}\mathbf{q}/(\omega/c)^2  }{\epsilon_0\bigl[(\omega+i\eta)^2 - c^2q^2\bigr]},
\end{equation}
with $\mathbf{U}$ representing the identity matrix.  This is the same as the usual dyadic Green's function up to a 
constant \cite{keller,novotny}. 

To obtain the free photon Green's function suitable for our geometry, we inverse Fourier transform $y$ and $z$ back to real
space and keep $q_x$ as it is.  
In our context, the free Green's function is a $2 \times 2$ matrix since the $y$ component is never needed. The explicit expression
for the $(x,z)$ block in atomic units ($4\pi \epsilon_0 = 1$) is:
\begin{equation}
	d^r(q_x,\omega,z,z^\prime) = \left[ \begin{array}{cc}
              \left(1 - \frac{c^2q_x^2}{\omega^2}\right)g & \frac{iq_xc^2k}{\omega^2}g_1 \\
		\\
		\frac{iq_xc^2k}{\omega^2}g_1 & g + \frac{c^2k^2}{\omega^2}g_2
	\end{array} \right],
\end{equation}
where $k = \sqrt{|\omega^2/c^2 - q_x^2|}$ is the wavevector component perpendicular to the $x$ direction, and $g$, $g_1$, and $g_2$ are functions involving modified Bessel functions $J$, $Y$, and $K$ \cite{bessel}. Specifically, if we define $X = kr$ and $r = |z - z^\prime|$, these functions are given by:

- For propagating waves:
\begin{align}
	g &= \frac{\pi}{c^2}\bigl[Y_0(X) - iJ_0(X)\bigr], \\
	g_1 &= \frac{\pi}{c^2}\bigl[-Y_1(X) + iJ_1(X)\bigr], \\
	g_2 &= -g - \frac{g_1}{X}.
\end{align}

- For evanescent waves:
\begin{align}
	g &= -\frac{2K_0(X)}{c^2}, \\
	g_1 &= \frac{2K_1(X)}{c^2}, \\
	g_2 &= g - \frac{g_1}{X}.
\end{align}

The expressions for $g_{(1,2)}$ depend on whether the waves are propagating ($\omega^2/c^2 > q_x^2$) or evanescent ($\omega^2/c^2 < q_x^2$). The modified Bessel functions $J_0$, $J_1$, $Y_0$, $Y_1$, $K_0$, and $K_1$ are used to describe the spatial dependence of the free photon Green's function in the $z$ direction. By substituting the appropriate expressions for $g$, $g_1$, and $g_2$ into the matrix for $d^r(q_x,\omega,z,z^\prime)$, we can handle both propagating and evanescent wave contributions to the photon Green's function within the system.

\section{Radiative heat transfer between coplanar 2D metals}

In the last section, we derived general formulas to calculate the RHT between coplanar 2D objects with a square lattice. However, the obtained tight-binding formulas Eqs.~(\ref{Pixx})-(\ref{Pizz2}) for the retarded photon self-energy $\Pi^r$ are complicated and time-consuming, necessitating further approximations for efficient calculations. First, because the thermal wavelengths of photons are much longer than those of electrons, we can neglect the wavevector dependence of the photon self-energy, i.e., we use the long-wavelength approximation ($q_x = 0$), which can significantly reduce the computational effort. This approximation is valid and widely adapted in studying RHT, especially for homogeneous materials \cite{zhu4,longwave}. Moreover, we assume that the two lattices are semi-infinite in the $z$ direction. With an increase in the separation distance $d$, a larger lattice cutoff $L_z$ value in the $z$ direction is needed to ensure convergence. As the value of $L_z$ determines the size of the self-energy matrices, the computational complexity grows rapidly, which becomes the major obstacle for actual calculations.

For pure metals, however, the electron behavior is well-characterized by the Drude model \cite{drude1,drude2}, which simplifies the expression for the retarded photon self-energy $\Pi^r$ as follows:
\begin{equation}
	\label{drude}
	\Pi^r_{\mu\nu}(\omega,l_z,l'_z)= \delta_{\mu\nu}\delta_{l_z,l'_z}\dfrac{a^2e^2\hbar\omega}{\hbar\omega+2i\eta}\int\dfrac{d{\bf q}}{4\pi^2}v^2_x\left( - \dfrac{{\rm d}f}{{\rm d}\epsilon} \right).
\end{equation}
Here, $v_x = 2at\,{\rm sin}(q_xa)/\hbar$, and $\epsilon = -2t\bigl[{\rm cos}(q_xa) + {\rm cos}(q_za)\bigr]$. It is diagonal in direction and local in sites. By applying Eq.~(\ref{drude}), not only can we circumvent the complex tight-binding formulas, but we can also handle a much larger effective lattice depth of $s\times aL_z$ by introducing a scale factor $s$. This is because the Drude model, lacking a characteristic length scale, renders the actual lattice constant $a$ irrelevant. Consequently, the converged value of $L_z$ can be significantly reduced for which the detailed comparison of these simplifications is provided in the supplementary materials \cite{supp}.

We applied the derived formulas to investigate the RHT between 2D metals using the following parameters and computational details. The hopping parameter $t$ is set to \SI{0.85}{\electronvolt}, and the damping parameter $\eta$ is \SI{27.2}{\milli\electronvolt}, which are typical values for simulating common metals \cite{llic}. The lattice dimensions are $L_z \times L=640\times640$, with a lattice constant $a$ of \SI{4}{\bohr} (4 times the Bohr radius), which is also the assumed thickness of the metal flake. For calculating photon self-energies, we used both the tight-binding method and the Drude model, with scale factor $s$ optimized to ensure convergence across various separation distances. To circumvent the divergence of the free photon Green's function when two electrons are at the same location ($r=0$), we impose a minimum distance cutoff $r_{\text{cut}}=\SI{1.6}{\bohr}$. The temperatures are maintained at $T_1 = \SI{1000}{\kelvin}$ and $T_2 = \SI{300}{\kelvin}$ with a null chemical potential. For comparison with black-body radiation, the heat transfer rate per unit length is also calculated using the Stefan-Boltzmann law: $J_{\text{bb}}=a F_{12}\sigma\, (T_1^4-T_2^4)$, where $F_{12}=\frac{a}{2d}$ is the geometrical view factor, and $\sigma \approx \SI{5.67e-8}{\watt\per\square\meter\per\kelvin\tothe{4}}$ is the Stefan-Boltzmann constant \cite{Hurtado1,Hurtado2}. 

\begin{figure}
	\includegraphics[width=8.6 cm]{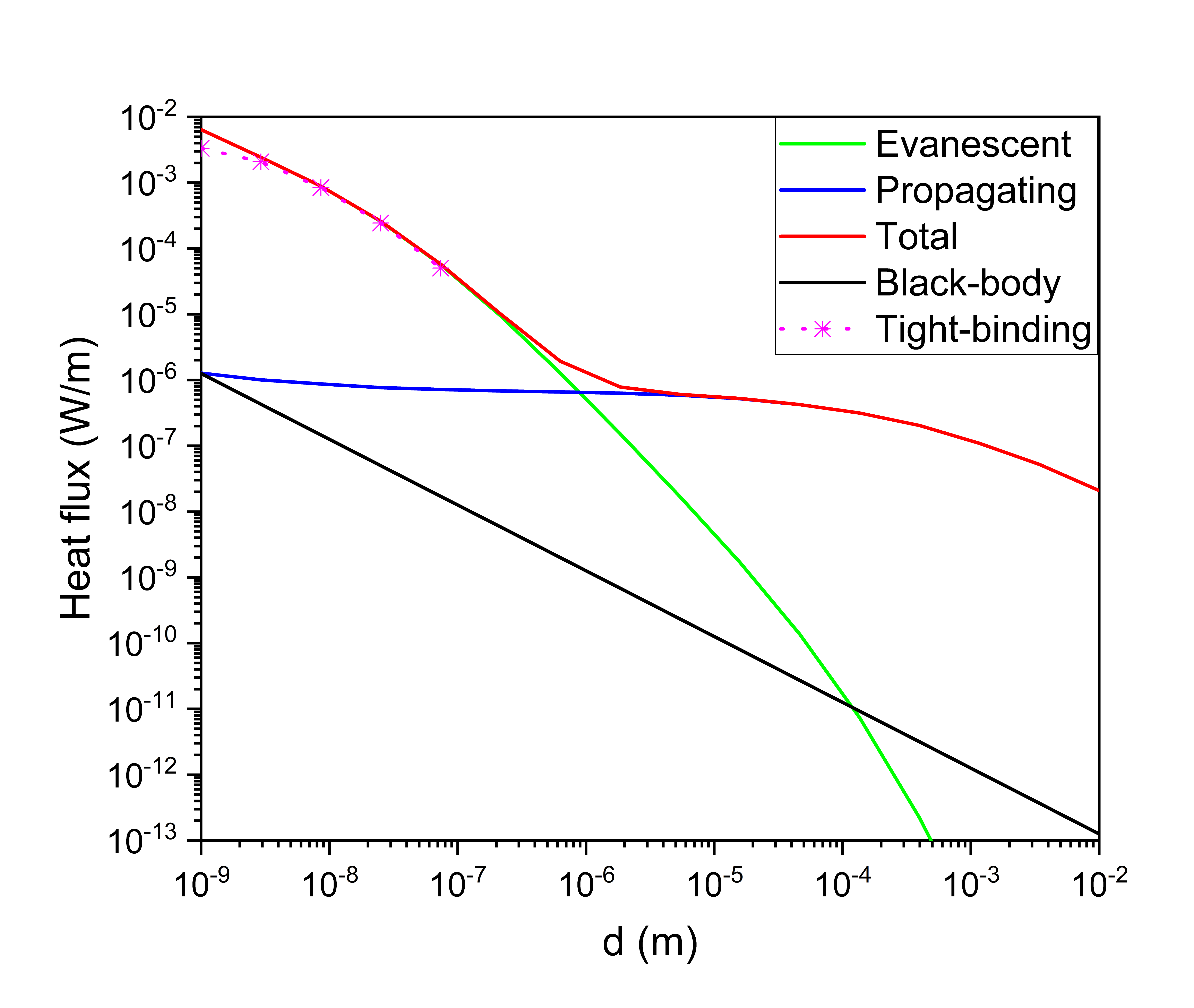}
	\caption{The distance dependence of radiative heat flux density between two coplanar metal sheets with temperatures $T_1 = \SI{1000}{\kelvin}$ and $T_2 = \SI{300}{\kelvin}$. The dashed curve with symbols corresponds to results from the tight-binding method, while the colored solid curve represents results from the Drude model. The black solid lines depict the heat flux density of black-body radiation calculated by the Stefan-Boltzmann law, factoring in the geometrical view factor $F_{12}=\frac{a}{2d}$.}
	\label{fig2}
\end{figure}

In Fig.~\ref{fig2}, we present the calculated heat flux density between two coplanar 2D metal sheets as a function of the gap size. Results obtained from the tight-binding method with the long-wavelength approximation represented by a short dashed line with symbols, are only converged for separation distances up to \SI{0.1}{\micro\meter}. For larger gaps, $L_z = 640$ proves insufficient, and extending it further exceeds our computational limits. The results from the tight-binding method and the Drude model display good agreement, with only minor deviations in the extreme near-field regime at nearly contacting distances. With the same parameters used (such as $L_z \times L=640\times640$), the agreement concurs in both near-field and far-field regimes, which is further detailed in the supplementary materials \cite{supp}. Therefore, subsequent discussions will focus on results from the Drude model.

As depicted in Fig.~\ref{fig2}, the RHT between coplanar 2D metals decreases monotonically with increasing distance. Nonetheless, it is substantially larger than the black-body radiation at all measured distances, indicating a global super-Planckian RHT. Unlike a constant black-body radiation for face-to-face planar geometry, the black-body heat flux in coplanar geometry has a $1/d$ dependence due to the geometrical view factor $F_{12}=\frac{a}{2d}$ \cite{Hurtado2}. In the extreme near field, around \SI{1}{\nano\meter}, the total heat flux density can reach up to $0.01 \,\text{W/m}$, nearly four orders of magnitude greater than that of black-body radiation, which is a typical characteristic of near-field RHT. Interestingly, a $1/d$ dependence is observed at short separations due to the long-wavelength approximation used in the calculations \cite{longwave}. Accounting for the full spatial dispersion would likely reveal a saturation trend in the extreme near field \cite{zhu1,zhu3}. The heat flux density decreases rapidly within the near field and then more gradually from \SI{1}{\micro\meter} to \SI{1}{\milli\meter}. Beyond $d = \SI{1}{\milli\meter}$, the heat flux begins to mirror the $1/d$ dependence of black-body radiation yet remains several orders of magnitude larger, in agreement with previous studies \cite{Hurtado2}.

In Fig.~\ref{fig2}, we also separated the contributions from evanescent and propagating waves to identify the tunneling channels. The varying decreasing trend in heat flux at different distances is attributed to the transition from evanescent to propagating waves. For $d<\SI{1}{\micro\meter}$, evanescent waves dominate the heat transfer, exhibiting rapid decay with distance, which is typical for near-field RHT. At a separation distance of approximately 100 nanometers, the calculated heat flux density is around $10^{-4}$ W/m, which is in good agreement with recent experimental measurements of 830 W/m$^2$/K for coplanar silicon carbide membranes \cite{tanglei}. As the distance exceeds \SI{1}{\micro\meter}, evanescent waves diminish and propagating waves gradually become the primary contributors. It should be noted that, in contrast to the face-to-face planar geometry, black-body radiation between coplanar objects decays as $1/d$ due to the view factor. However, the heat flux between coplanar 2D metals from propagating waves in the extreme near field is comparable to that of black-body radiation. It remains nearly constant up to $d=\SI{100}{\micro\meter}$, resulting in super-Planckian behavior even in the far field. Notably, for $d>\SI{1}{\milli\meter}$, the propagating heat flux again follows a $1/d$ dependence similar to black-body radiation but maintains a magnitude millions of times larger, consistent with prior observations. This suggests that the RHT facilitated by propagating waves in coplanar configurations exhibits characteristics similar to traditional face-to-face geometry, where propagating heat flux remains constant over distance \cite{song}.

\begin{figure}
	\includegraphics[width=8.6 cm]{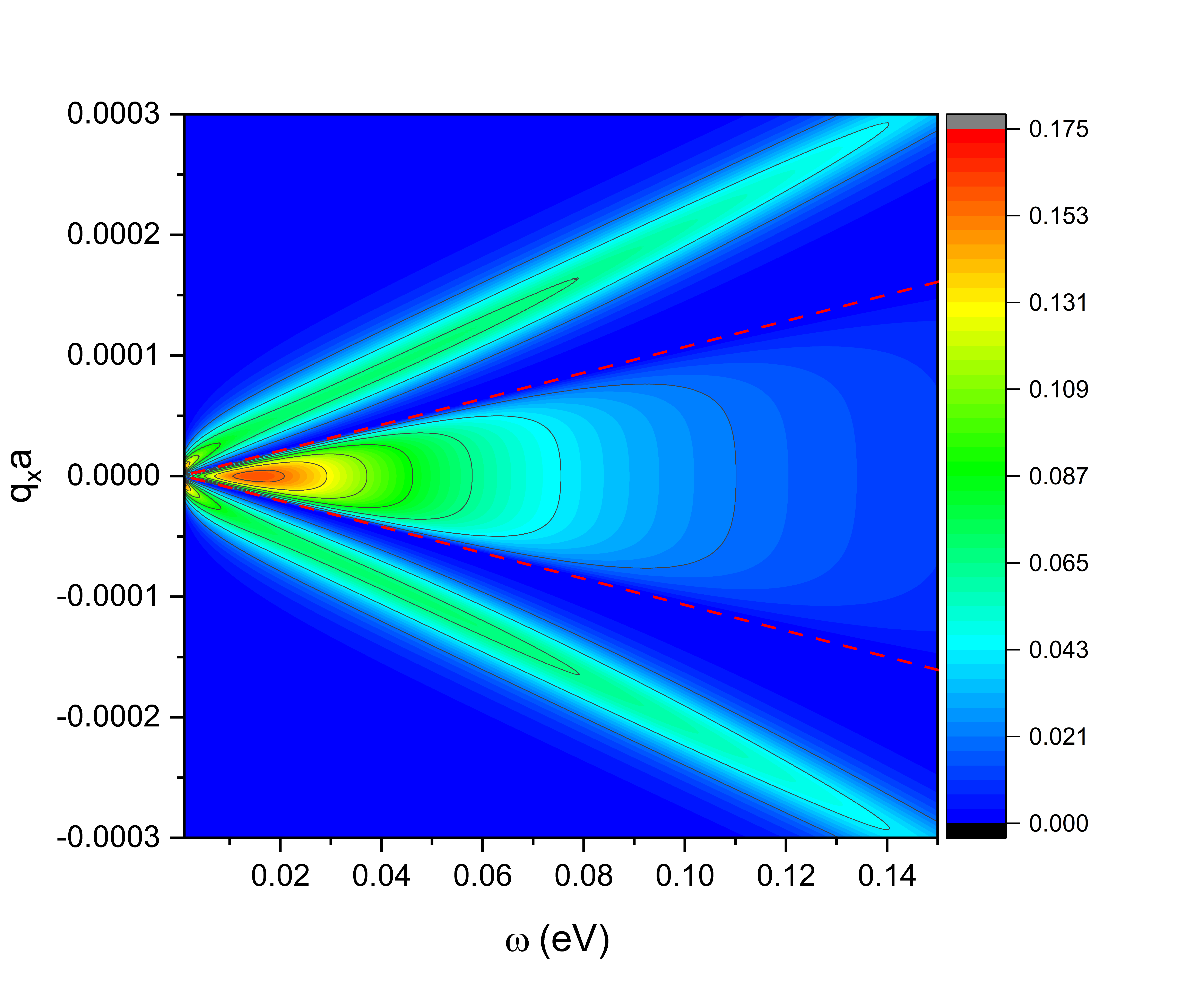}
	\caption{The spectrum of transmission function between two coplanar metal sheets with the gap size of $\SI{1}{\micro\meter}$. The horizontal coordinate is the frequency, and the vertical coordinate is $q_x\times a$. The temperature is fixed at $T_1=$ 1000 K and $T_2=$ 300 K. The red dashed lines represent the light line $q^2 = \omega^2/c^2$, the boundary between evanescent and propagating modes.}
	\label{fig3}
\end{figure}

The dimensionless spectrum transmission function $T(q_x,\omega)$ between coplanar 2D metals was analyzed and is presented in Fig.~\ref{fig3}. The gap size was fixed at \SI{1}{\micro\meter}, a distance at which both evanescent and propagating waves significantly contribute to RHT. The heatmap depicted in Fig.~\ref{fig3} is distinctly divided into two regions. The central region, or the ``body" of the heatmap, corresponds to the spectrum of propagating waves, which are confined by the relation $q^2 < \omega^2/c^2$. Most notable is that the majority of contributions within this region are from low-frequency modes ($<\SI{0.1}{\electronvolt}$) which align with the thermal energy range defined by the temperatures of the bodies involved (1000 K and 300 K). Moreover, we observe contributions from evanescent waves on the ``wings" of the heatmap. These contributions are predominantly concentrated within a narrow band with a dispersion relation close to $\pm\omega/c$. This suggests that the evanescent waves may stem from the coupling of surface plasmon resonance, which typically occurs at these higher frequency multiples relative to the light line ($q^2 = \omega^2/c^2$) \cite{joulain}. The suspected significant role of surface plasmons in near-field RHT highlights the unique mechanisms of heat transfer at the nanoscale, distinguishing it from the classical radiative heat transfer observed at larger scales.
\section{Conclusion}

In summary, we have proposed a fully quantum-mechanical microscopic theoretical framework to study radiative heat transfer between coplanar objects and systematically applied it to the two-dimensional metal configurations. By employing both tight-binding and Drude models within the NEGF formalism, our investigations reveal that the radiative heat transfer in these sub-wavelength systems significantly exceeds that of black body radiation across all distances, establishing a regime of global super-Planckian RHT. The distance dependence of the heat flux density is characterized by a rapid attenuation in the near field, transitioning to a more gradual reduction in the far field, and ultimately conforming to a $1/d$ behavior, while maintaining a substantially higher magnitude than that predicted by black-body radiation.

The analysis of the tunneling channel and dispersion relation of the spectral function has elucidated the distinct contributions from both evanescent and propagating waves to the RHT. Evanescent waves are identified as the primary heat transfer mechanism at shorter separations, whereas propagating waves dominate as the separation increases. Remarkably, the heat flux associated with propagating waves exhibits an almost invariant behavior over an extended range of distances up to \SI{100}{\micro\meter}, indicative of the super-Planckian transport in the far field. The dispersion relation of the spectrum function between coplanar 2D metals reveals the distinct contributions from propagating and evanescent waves. Most of the propagating wave contributions coming from low frequencies and wavevectors while evanescent wave contributions being limited to a narrow range close to the light cone, possibly originating from surface plasmon resonance. 

This work advances the fundamental understanding of radiative heat transfer in nanostructured systems, particularly emphasizing the potential of coplanar 2D objects to enable heat transfer that surpasses classical limits. The derived tight-binding formulas are universally applicable to coplanar 2D square lattices, while the Drude model is limited to pure metals. Future research could expand upon this study by investigating different materials and geometric configurations, as well as by incorporating the effect of electron conduction, which becomes dominant at extremely small distances. The findings presented herein paves the way for enhanced thermal management and energy efficiency in nanoscale devices.

\section*{Acknowledgments}
T.Z. is supported by National Natural Science Foundation of China (Grant No. 12204346). J.-S. W acknowledges support from the Ministry of Education, Singapore, under the Academic Research Fund Tier 1 (A-8000990-00-00). 

\bibliographystyle{apsrev4-2}
\bibliography{Super-Planckian.bib}
\end{document}


\title{Supplementary materials for ``Super-Planckian radiative heat transfer between coplanar two-dimensional metals"}
\author{Tao Zhu}
\email{zhutao@tiangong.edu.cn; phyzht@outlook.com}
\affiliation{School of Electronic and Information Engineering, Tiangong University, Tianjin 300387, People's Republic of China}
\author{Yong-Mei Zhang}
\affiliation{College of Science, Nanjing University of Aeronautics and Astronautics, Jiangsu 210016, People's Republic of China}
\author{Jian-Sheng Wang}
\affiliation{Department of Physics, National University of Singapore, Singapore 117551, Republic of Singapore}
\date{\today}
\maketitle
\begin{widetext}
	\setcounter{equation}{0}
	\setcounter{figure}{0}
	\setcounter{table}{0}
	\makeatletter
	\renewcommand{\theequation}{S\arabic{equation}}
	\renewcommand{\thefigure}{S\arabic{figure}}
\begin{figure}
	\includegraphics[width=0.9\columnwidth]{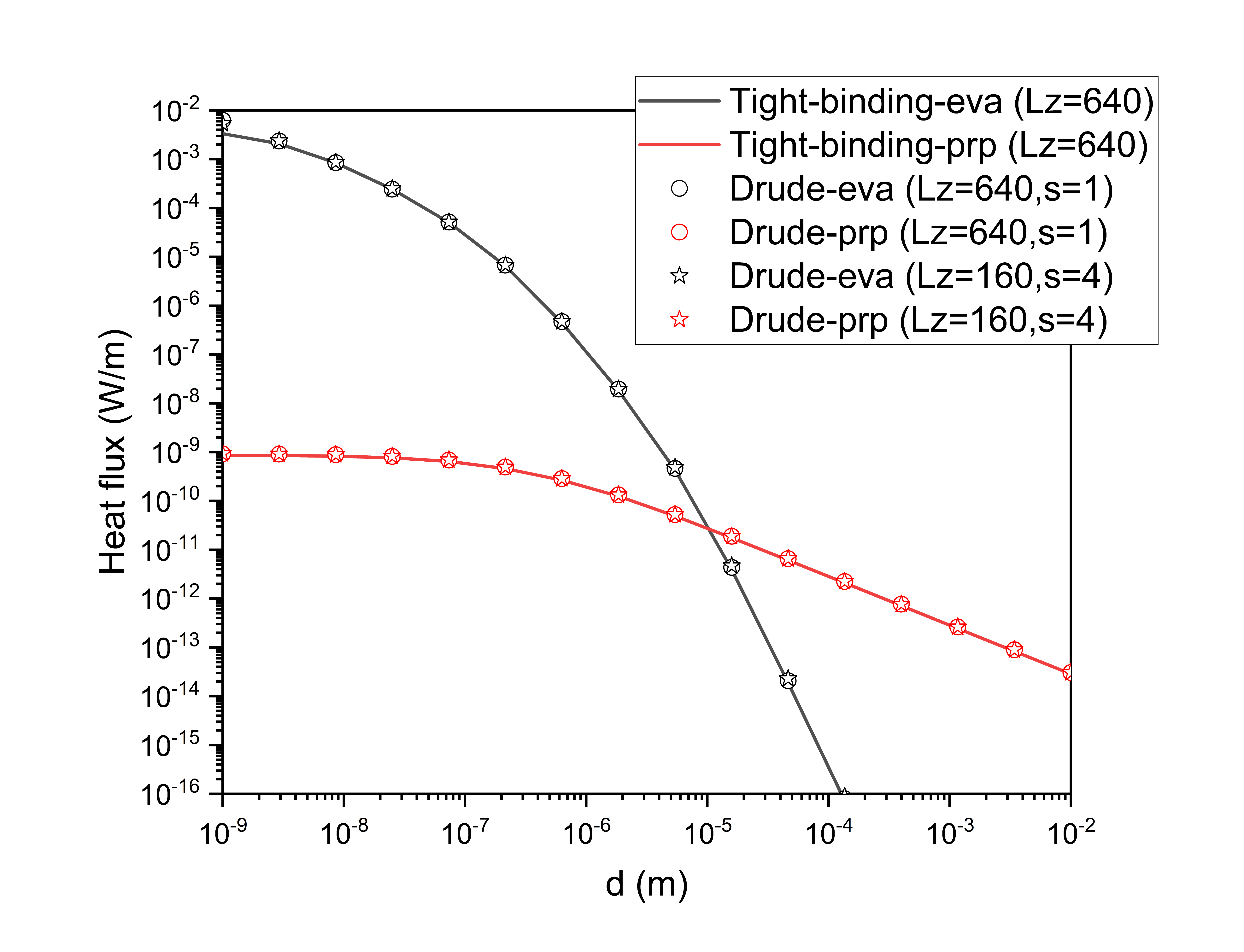}
	\caption{The distance dependence of radiative heat flux density between two coplanar metal sheets with temperatures $T_1 = \SI{1000}{\kelvin}$ and $T_2 = \SI{300}{\kelvin}$. The ``eva" and ``prp" represent contributions from evanescent waves and propagating waves, respectively. The solid lines correspond to results from the tight-binding method with $Lz=640$, the circle symbol represents results from the Drude model with $Lz=640$, and the star symbols are results from the Drude model with $Lz=160$ but the lattice in $z$ direction is scaled by a factor $s = 4$.}
	\label{figs1}
\end{figure}
We show in Fig. \ref{figs1} the comparison between heat flux calculated from the tight-binding model, the Drude model, and the Drude model with a scale factor. As pointed out in the main text, although $L_z=640$ is insufficient to ensure a full convergence, it can be seen that results from the tight-binding model and the Drude model agree at all distances with the same parameter. On the other hand, by adopting the lattice scale factor $s$, we can use a smaller value of $L_z=160$ with $s=4$ to obtain the same results as not scaled Drude model with $L_z=640$, which significantly saved the computational efforts. This agreement indicates that our approximations used in the main text is valid.
\end{widetext}